\documentclass[twocolumn,preprintnumbers,amsmath,amssymb,aps]{revtex4}

\usepackage{graphicx}	
\usepackage{dcolumn}	
\usepackage{bm}		
\usepackage{color}
\usepackage{colordvi}
\usepackage{epsfig}

\newcommand{\kap}{\boldsymbol{\kappa}}

\newcommand{\qb}{{\bf q}}
\newcommand{\kb}{{\bf k}}
\newcommand{\pb}{{\bf p}}
\newcommand{\sig}{\boldsymbol{\sigma}}

\newcommand{\B}{\boldsymbol{B}}

\newcommand{\Q}{\boldsymbol{Q}}
\newcommand{\Finger}{\boldsymbol{B}}
\newcommand{\E}{\boldsymbol{E}}

\begin{document}

\title{Glass rheology: From mode-coupling theory to a dynamical yield criterion}

\author{Joseph M. Brader$^{1}$, Thomas Voigtmann$^{1,2}$, Matthias Fuchs$^{1}$, Ronald G. Larson$^{3}$ 
and Michael E. Cates$^{4}$}
\affiliation{$^{1}$Fachbereich Physik, Universit\"at Konstanz, D-78457 Konstanz, Germany.\\ 
$^{2}$Institut f\"ur Materialphysik im Weltraum, Deutsches Zentrum f\"ur Luft- und Raumfahrt (DLR), 51170 K\"oln, Germany and Zukunftskolleg der Universität Konstanz.\\ 
$^{3}$Department of Chemical Engineering, University of Michigan, Ann Arbor, Michigan 48109-2136, USA.\\
$^{4}$SUPA, School of Physics and Astronomy, University of Edinburgh, Kings Buildings, Mayfield Road, Edinburgh EH9 3JZ, 
Scotland.
}

\date{\today}

\begin{abstract}
  The mode coupling theory (MCT) of glasses, while offering an incomplete description of glass transition physics, represents the only established route to first-principles prediction of rheological behaviour in nonergodic materials such as colloidal glasses. However, the constitutive equations derivable from MCT are somewhat intractable, hindering their practical use, and also their interpretation. Here we present a schematic (single-mode) MCT model which incorporates the tensorial structure of the full theory. Using it, we calculate the dynamic yield surface 
for a large class of flows. 
\end{abstract}

\maketitle

The 20th Century saw formidable advances in the subject known as theoretical rheology -- whose aim is to predict or explain the nonlinear flow behavior of materials. Ideally for each class of material one wishes to gain a `constitutive equation' which predicts the stress tensor at time $t$ as a functional of the strain tensor at all earlier times (or vice versa). There are two broad approaches to this task. The more traditional one focuses on symmetry, conservation, and invariance principles (often of some subtlety) and then proposes empirical equations that respect these principles \cite{noll}. In the second approach, the goal
is to start from a first-principles analysis of molecular motion, and then by judicious (though
possibly uncontrolled) approximation arrive at a continuum-level constitutive model. This is clearly
far more ambitious, and success has so far been restricted to relatively few classes of material.
Perhaps the most striking success has been the Doi-Edwards theory for solutions and melts of
entangled linear polymers \cite{doi_edwards,larson} (extended later to branched \cite{tomstuff} or
breakable \cite{catesfielding} chains). In their resting state, such polymers are ergodic and therefore attain the Boltzmann distribution: moreover their local structure is weakly perturbed from this, even under flow.

Glasses at rest, in contrast, are nonergodic on experimental timescales. This poses major obstacles to the rheological theory of glasses, and is responsible for aging and other phenomena that have been partially addressed using mesoscopic models \cite{fielding}. The onset of arrest at the glass transition is, familiarly, accompanied by the onset of an elastic modulus. Window glass is a brittle solid: it deforms elastically for low stresses but shatters under large ones. However, some other glasses -- most notably in colloidal suspensions (whose glass transition, for hard spheres, is found experimentally at about 58\% volume fraction) are not brittle solids but show continuous yielding behavior. Although experiments suggest a more complex picture \cite{besseling,isa}, the simplest explanation is that, above some yield stress, the glass melts. If a steady stress above the yield level is maintained, the resulting fluid can be expected to attain an ergodic (though non-Boltzmann) steady state. 

This restoration of ergodicity under steady flow offers one motivation for an approach to glass rheology based on mode-coupling theory (MCT). In particular, it mitigates a well-known shortcoming of MCT which can address only the approach to the glass transition from the liquid side, and therefore cannot access aging phenomena. For systems at rest MCT predicts a true glass transition (that is, a divergent relaxation time) rather than one cut off by activated processes. MCT does not address such processes, at least in its simplest form (for a discussion of extensions that do, see \cite{catesramaswamy,chong}); on the other hand, despite its mean-field character, MCT does appear to capture some aspects of dynamic heterogeneity \cite{biroli}. 
When applied to colloid rheology, MCT addresses a set of Brownian particles advected by a flow that interact solely by conservative interparticle forces. As such it neglects hydrodynamic interactions whose role near the glass transition is unclear. 
(Such interactions are implicated in shear thickening, which can also arise in dense colloids; 
see e.g. \cite{holmes}.)

The application of MCT to systems under flow requires additional approximations beyond those of
quiescent MCT. The most comprehensive route involves an `integration through transients' (ITT) formalism which 
has been detailed in a series of papers addressing in turn steady shear \cite{catesfuchs}, time-dependent shear \cite{brader1} and arbitrary homogeneous, time-dependent (incompressible) flows \cite{brader2}. 
Below we very briefly outline the final constitutive model of \cite{brader2}, before turning to our main purpose in the present work. This is to present a radically simplified model, directly inspired by those results, but far more suitable for practical calculations of flow behavior. It is derived, in essence, by suppressing all wavevector indices on the MCT equations for coupled density fluctuations, resulting in a single-mode description. Our model thus stands in relation to MCT-ITT as the well known `schematic' models of the MCT glass transition stand in relation to quiescent MCT \cite{goetze}. It builds on previous schematic models for MCT rheology \cite{faraday,catesfuchs} which were however restricted to steady shearing, and contain no tensorial information about the stress response to more general flows. In the present work we overcome both of these limitations, gaining a practicable but still microscopically motivated constitutive description for glass rheology.

In the MCT-ITT approach, the system is initially assumed to be at rest and in Boltzmann equilibrium. The flow is subsequently switched on; its effect via particle advection is to create a nonequilibrium Smolochowski operator. An exact nonequilibrium Green-Kubo relation allows the stress tensor at any later time to then be expressed as a time integral of the expectation value of a certain operator evaluated within the equilibrium ensemble \cite{faraday,fuchslong}. Applying MCT-type factorizations to this expectation value gives a constitutive description in three parts. The first relates the deviatoric stress to strain history via a time integral whose (wavevector dependent) kernel involves derivatives (with respect to wavenumber $k$) of the 
equilibrium static structure factor $S_k$ and the square of a normalized transient density 
auto-correlator:
\begin{eqnarray}
\sig(t) &=& -\int_{-\infty}^{t} \!\!\!\!\!\!dt'\!\int\!\!\!
\frac{d{\bf k}}{32\pi^3} 
\left[
\frac{\partial}{\partial t'}(\kb\!\cdot\!\Finger(t,t')\!\cdot\!\kb)
\,\kb\kb\right]\times \label{nonlinear}
\\
&\times&
\left[
\left(\!\frac{S'_k S'_{k(t,t')}}{k\,k(t,t')S^2_k}\!\right)\Phi_{{\bf k}(t,t')}^2(t,t')
\right].\notag
\end{eqnarray}
This expression is found from results reported in \cite{brader2} (suppressing an isotropic pressure 
term) and is a consequence of approximating the relaxation of stress fluctuations by 
that of the density fluctuations responsible for slow stuctural relaxation in dense systems.
$\Phi_{\bf k}(t,t')$ is a transient density correlator which, by virtue of the MCT-ITT approach, is calculated using the equilibrium distribution. The correlator measures the overlap of a density fluctuation at wavevector ${\bf k}$ and time $t$ with one at earlier time $t'$ whose wavevector $\bar{\bf k}(t,t')$ evolves due to flow-induced advection to become ${\bf k}$ at $t$. For the purposes of calculating this correlator, the system is taken to be in equilibrium at $t'$, with the flow acting thereafter. 
Because averaging in MCT-ITT can be done with the equilibrium distribution function, only the equilibrium structure factor 
$S_k$ appears in Eq.1.
In Eq.\ref{nonlinear}, $\B$ is the Finger tensor: 
\begin{eqnarray}
\Finger(t,t')=\E(t,t')\cdot \E^T(t,t'),
\end{eqnarray}
where ${\E}(t,t')$ is the deformation tensor applied between a past time $t'$ and the present time $t$. 
(This obeys $\partial \ln{\E}/\partial t = \kap(t)$, where $\kap$ is the strain-rate tensor.) 
The reverse-advected wavevector is given as 
${\bf k}(t,t')={\bf k}\cdot\E(t,t')$.
(Note that, where these differ, we use the notation of \cite{doi_edwards} rather than 
\cite{larson}. Note also that we follow the notation and formalism of \cite{brader2} which is modified from that of \cite{catesfuchs,brader1,faraday,fuchslong} as explained in detail in \cite{catesfuchs2}.)

The second part of the MCT-ITT description determines the time evolution of the correlators appearing in Eq.\ref{nonlinear} in terms of a three-time memory kernel:
\begin{eqnarray}
\dot\Phi_{\bf q}(t,t_0)
&+& \Gamma_{\bf q}(t,t_0)\bigg(
\Phi_{\qb}(t,t_0)
\label{equom}
\\
&+&
\int_{t_0}^t dt' m_{\qb}(t,t',t_0) \dot\Phi_{\qb}(t',t_0)
\bigg) =0
\notag
\end{eqnarray}
where the overdots denote partial differentiation with respect to the first time argument.
Here the `initial decay rate' obeys
$\Gamma_{\bf q}(t,t_0)=D_0\bar{q}^2(t,t_0)/S_{\bar{q}(t,t_0)}$ with $D_0$
a bare diffusivity, and $\bar{{\bf q}}(t,t') = {\bf q}\cdot\E^{-1}(t,t')$. 
Finally, MCT-ITT approximates the memory kernel $m_{\bf q}(t,t',t_0)$ by the factorized expression
\begin{eqnarray}
\label{approxmemory}
&& \hspace*{-1cm}
m_{\qb}(t,t'\!,t_0) \!\!= \!\!
\frac{\rho}{16\pi^3} \!\!\int \!\! d\kb
\frac{S_{\bar{q}(t,t_0)} S_{\bar{k}(t',t_0)} S_{\bar{p}(t',t_0)} }
{\bar{q}^2(t',t_0) \bar{q}^2(t,t_0)}\\
&\times&
V_{\qb\kb\pb}(t',t_0)\,V_{\qb\kb\pb}(t,t_0)\Phi_{\bar{\kb}(t',t_0)}(t,t')
\Phi_{\bar{\pb}(t',t_0)}(t,t'),
\notag
\end{eqnarray}
where $\pb=\qb-\kb$, and the vertex function obeys
\begin{eqnarray}
\!\!\!\!\!V_{\qb\kb\pb}(t,t_0) \!=\! \bar\qb(t,t_0)\cdot(
\bar\kb(t,t_0) c_{\bar{k}(t,t_0)} \!+
\bar\pb(t,t_0) c_{\bar{p}(t,t_0)}),
\label{vertex}
\end{eqnarray}
with $c_k = 1-1/S_k$. Again, like in Eq.\ref{nonlinear}, this assumes that fluctuating 
stresses decay mainly by density fluctuations.
In Eqs.\ref{nonlinear}--\ref{vertex}, MCT-ITT offers a closed constitutive 
model for interacting Brownian particles near the glass transition, requiring only 
$S_k$ and $\kap(t)$ as input to calculating the stress. 

So far, however, these equations have proved amenable only to approximate solution even for the simplest flows, such as steady and stepwise deformations \cite{catesfuchs,brader1,brader2}. In three dimensions, at least, such approximations have all involved replacing the complicated angular dependendence of correlators in ${\bf k}$-space by an isotropic dependence, creating the so-called
isotropically sheared hard sphere model or `ISHSM' \cite{faraday}.
(Avoidance of this has however very recently proved possible for two-dimensional systems \cite{oliver2D}.) 
Our aim here is to create a much more tractable, simplified description that nonetheless captures both the tensorial character and the basic time-dependence of the full MCT-ITT equations. In doing this we pay due attention to invariance properties which are respected by the microscopic theory but which could get lost in approximation. Of these, the least obvious is the 
principle of material objectivity, which requires invariance of the constitutive model under arbitrary time-dependent rotations. This stems from neglect of inertia (implicit in our description based on Brownian particles) and imposes strong constraints on the tensorial structure of rheological models \cite{noll}. 

\subsection{Derivation of schematic model}
Our starting point is Eq.\ref{nonlinear} for the stress tensor; to create a single-mode description we discard all wavevector-dependent information. 
A first step is to discard all angular information. 
To do so we assume that both the correlator $\Phi$ and the advected wavenumber $k(t,t')$ are isotropic in ${\bf k}$, so the 
integrand in Eq.\ref{nonlinear} becomes a product of an anisotropic and an isotropic term (each enclosed in square brackets). To see that this factorization respects the principle 
of material objectivity, note that the advected 
wavenumber $k(t,t')$ is not altered by a time dependent rotation $\Q(t)$, while the Finger tensor 
in the rotated frame is given by
$\hat{\Finger}(t,t')=\Q(t)\Finger(t,t')\Q^T(t)$.
Substitution into Eq.\ref{nonlinear} and changing 
integration variable to ${\bf k}'={\bf k}\cdot\Q(t)$ yields 
$\hat{\sig}(t)=\Q(t)\,\sig(t)\Q^T(t)$ as desired. 

With the isotropic assumption we can perform the angular integral in Eq.\ref{nonlinear} explicitly, to obtain 
\begin{eqnarray}
\sig(t) &=& \int_{-\infty}^{t} \!\!\!\!\!\!dt'\! 
\left[
-\frac{\partial}{\partial t'}
\Finger(t,t')
\right] G(t,t')\,,\label{Gform}\\
G(t,t') &=&
\frac{1}{60\pi^2}
\int dk\,k^5
\left(\!\frac{S'_k S'_{k(t,t')}}{k(t,t')S^2_k}\!\right)\Phi_{k(t,t')}^2(t,t')\,.
\notag
\end{eqnarray}
Having dealt explicitly with the tensorial structure, we may now safely discard all wavevector dependence in the expression for the generalized modulus $G(t,t')$. Thus we obtain a fully schematic model in which Eq.\ref{Gform} holds with
\begin{eqnarray}
G(t,t') = v_{\sigma}
\Phi^2(t,t')
\label{schematicT2}
\end{eqnarray}
where $v_{\sigma}= G(t,t)$ is a parameter measuring the strength of stress fluctuations. 
When analyzing flow curves of hard sphere-like colloids values of the order $100 k_BT/d^3$ are 
typically obtained, where $d$ is the sphere diameter \cite{fuchs3}.

The full microscopic equation of motion for the correlators is given by Eqs.\ref{equom}--\ref{vertex}. Discarding all wavevector indices in Eq.\ref{equom} gives the schematic representation
\begin{eqnarray}
\dot\Phi(t,t_0) 
+ \Gamma\bigg(
\Phi(t,t_0) + \int_{t_0}^{t}dt'\, m(t,t',t_0)
\dot\Phi(t',t_0)
\bigg) =0,
\label{eom2}
\end{eqnarray}
There is now only one initial decay rate, $\Gamma$, which sets the microscopic time scale 
and may thus be set equal to unity. 

The microscopic form for the memory function (Eq.\ref{approxmemory})  depends quadratically on the 
correlators and has vertices 
(Eq.\ref{vertex}) that are a function of the accumulated strain over different time intervals. 
This strain dependence of the memory kernel is a central novel aspect of our microscopic constitutive equation 
(Eqs.\ref{nonlinear}--\ref{vertex}) the qualitative aspects of which we aim to reproduce in a 
simplified schematic model.
In the absence of flow, our memory function reverts to that of standard (quiescent) MCT, with a 
time-independent coupling to a product of two stationary correlators. 
The arrest transition of this 
standard MCT is captured by a minimal schematic model (replacing in effect 
the ${\bf k}$ integral by a single wavevector) known as the $F_{12}$ model \cite{goetze}, 
with both a linear and a quadratic coupling. 
It is well established that the schematic $F_{12}$ model 
captures quantitatively many universal aspects of the glass transition contained in the full MCT
equations.
To extend the $F_{12}$ model to address rheology we now make the ansatz
\begin{eqnarray}
m(t,t',t_0)=h_1(t,t_0)\,h_2(t,t')\,h_3(t',t_0) \times
\notag
\\
\times[v_1\Phi(t,t')+v_2\Phi^2(t,t')]. \notag
\end{eqnarray} 
whose form is inspired directly by Eq.\ref{approxmemory}. So long as the product of the three $h$ factors becomes constant in the absence of flow, the 
standard $F_{12}$ model is recovered in that limit. 

Within the microscopic description, the wavelengths of density fluctuations are on average 
reduced by deformation, allowing diffusion and interparticle forces to relax these fluctuations more rapidly, 
so that strain causes a progressive loss of memory 
(this mechanism is represented by the time dependence of the vertex functions 
in Eq.4). The $h_i$ should thus be decaying functions of the 
strain accumulated between their two time arguments. However, there is some redundancy between these factors and we have found that if $h_1$ and $h_2$ both 
decay, $h_3$ can be set to unity without 
losing any obvious physical content of the model.
In addition we set $h_1=h_2=h$ so that 
the model contains only one unknown function $h(t_1,t_2)$; choosing this as detailed below gives good qualitative agreement with the full theory for all flows so far investigated. By these considerations we arrive at the following expression for the schematic memory function:
\begin{eqnarray}
m(t,t',t_0)=h(t,t_0)\,h(t,t')\,[\,v_1\Phi(t,t')+v_2\Phi^2(t,t')\,].
\label{memory}
\end{eqnarray} 

Our final task is to decide the form of $h(t_1,t_2)$. This is shorthand for $h(\E(t_1,t_2))$, and encodes the loss of memory caused by strain. 
Earlier work on steady shear shows that the choice $h = 1/(1+(\gamma/\gamma_c)^2)$ 
(with $\gamma = (t_2-t_1)\dot\gamma$ the accumulated shear strain) gives results close to the full MCT-ITT theory for this case \cite{faraday}. We can be guided by this finding, but need to generalize it to non-shear flows, such as elongational deformation, which can be expected to differ quantitatively in their effects on memory loss. Notably, this variation is strictly limited in form: for incompressible systems, the principle of material objectivity ensures that the deformation tensor ${\E}$ can enter only through the invariants $I_1 = {\rm Tr} \Finger$ and $I_2 = ({\rm Tr}\Finger^{-1})$ \cite{larson}. 
We are thus led to the following choice
\begin{eqnarray}
h(t,t_0)=\frac{\gamma_c^2}{\gamma_c^2 + 
\left[\nu I_1(t,t_0) + (1-\nu)I_2(t,t_0)-3\right]}\,,
\label{modified_h}
\end{eqnarray} 
where we have introduced both a mixing parameter $\nu$ ($0 \le \nu \le 1$) and 
a crossover strain parameter $\gamma_c$. 
(The latter sets the scale for the recoverable elastic strain; 
with this as a fit parameter, schematic models accurately account for structural distortions 
in steady state shear \cite{fuchs3}.)
Note that $I_1=I_2$ for both shear and planar extensional flows
whereas $I_1\neq I_2$ for uniaxial elongation. 

Eqs.\ref{Gform}--\ref{modified_h} fully specify our new schematic model.
In a further (conventional) simplification, the parameters $v_1$ and $v_2$ are now replaced by
$v_2=2$ and $v_1=2(\sqrt{2} -1) + \varepsilon/(\sqrt{2}-1)$. The `separation parameter' 
$\varepsilon$ is analogous to the density in a microscopic system, with negative values corresponding 
to fluid states and positive values corresponding to glass states. 
We have thus obtained a closed constitutive model with four adjustable parameters 
($v_{\sigma}, \Gamma, \gamma_c$ and $\nu$) and one control parameter ($\epsilon$). 
This compares with the full MCT-ITT equations in which the static structure factor $S_q$, density and bare diffusivity control all measurable quantities. Accordingly one can view $v_{\sigma}, \gamma_c$ and $\nu$ as fixed in some unspecified way by $S_q$. 
The latter in turn depends on the thermodynamic control parameters and on the interaction potential between particles. 
When applied to experimental data, all five parameters of the model are used to fit the data and, as expected, exhibit smooth 
variation with the external parameters 
\cite{fuchs3}.

\subsection{Physical content of the model}
The constitutive model laid out above describes a fluid whose {\em instantaneous} elastic
response is linear in the Finger tensor $\Finger$, as manifest in Eq.\ref{Gform}. 
The Finger tensor is a natural frame-invariant extension of the linear strain tensor, and 
a similar dependence of stress on strain arises, for instance, in models that involve networks 
of Hookean springs. 
Indeed, were $G(t,t')$ in Eq.\ref{Gform} to be replaced by a strain-independent,
time-translation-invariant kernel $G(t-t')$, this would recover the Lodge equation \cite{noll,larson}, 
a standard rheological model. (The relationship to such models is explored further in the Supplementary Information, SI.)
The additional nonlinearity in our model arises almost entirely from the strain-induced 
erasure of memory. 
This differs somewhat from Eq.\ref{nonlinear} in which there is some additional nonlinear 
elasticity, even setting $\Phi=1$; but this does not survive the schematic-model relaxation 
approximation that gives Eq.\ref{schematicT2}.

Thus the kernel $G(t,t')$ in Eq.\ref{schematicT2} is strongly sensitive to the strain-induced decay of the correlators, as found in turn via Eq.\ref{equom}. This decay sets in at strain increments of order $\gamma_c$ and is absent in the linear response regime for which $G(t-t') = v_\sigma\Phi_e(t-t')^2$ with $\Phi_e$ the quiescent state correlator. For $\epsilon < 0$ linear response describes a viscoelastic fluid; for $\epsilon >0$, a viscoelastic solid. In the latter (glass) regime,  a sustained strain rate causes the memory (and therefore the correlators) to fall to zero at long times, resulting in fluidization. When such a flow ceases, the system solidifies again. Because of the memory-function structure, the system's current material properties (for example, the relaxation time that governs response to an additional strain increment) are influenced by past deformations; but this memory is itself erased by large strains. Thus, for instance, if a steady flow is suddenly switched off, it takes some time for the memory kernel to rebuild, the correlators to slow down, and the solid-like properties of the glass to be reinstated. 

Note that the model captures memory erasure by sudden (step-strain) events as well as by sustained deformation rates. Here there are considerable subtleties to the model. For instance $m(t,t',t_0)$ is diminished by a step strain that occurs at any time between $t_0$ and $t$, as makes sense given that the time integral in Eq.\ref{eom2} spans this interval. However, if this step strain is later reversed, only pairs of steps that bracket the intermediate time $t'$ in that integral contribute to a reduction of memory. Such features are sharpened by taking the schematic limit, but so far as we can tell they do have proper anticedents within the full model, 
Eqs.\ref{equom}--\ref{vertex}.  

\if{
The Lodge equation is a member of the K-BKZ class of equations \cite{larson}, for which
\begin{eqnarray}
\sig(t)=2\int_{-\infty}^{t}\!\!dt'\;\left(
\frac{\partial u}{\partial I_1}\Finger(t,t') - \frac{\partial u}{\partial I_2}\Finger^{-1}(t,t')
\right).
\label{kbkz}
\end{eqnarray}
with $u(I_1,I_2,t-t')$ a function whose time integral gives the strain energy. Even allowing for breakdown of time translation invariance, so that $u = u(I_1,I_2,t,t')$, we find that our model is not of this form (except for the limiting case of $\nu = 1$). However, modulo the same generalization of the time arguments, our model does belong to an extension of the Rivlin-Sawyers class of equations \cite{hassager} in which the coefficients of $\Finger$ and $\Finger^{-1}$ in Eq.\ref{kbkz} 
need not satisfy an integral relationship. 
Note that the generalization to two time arguments in the strain energy kernel $u$
can give rise to a stress response significantly different from that of the standard K-BKZ theory.
Casting our model into differential form sheds further light on its relationship to the upper-convected 
Maxwell equation (see supporting information (SI)).
}\fi

\begin{figure}
\includegraphics[width=8cm]{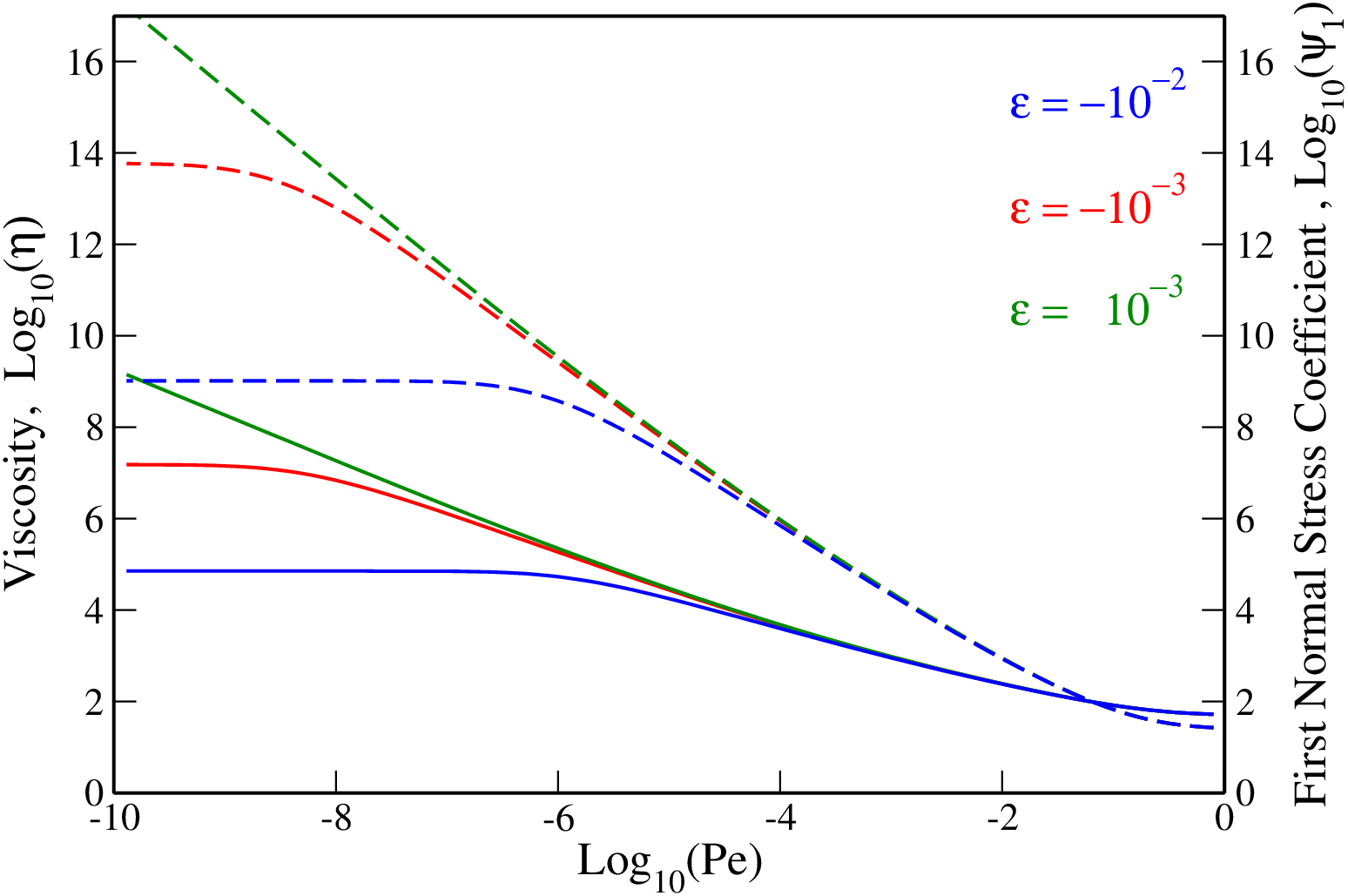}
\caption{
The steady state shear viscosity $\eta$ (full lines) and first normal stress coefficient 
$\Psi_1=(\sigma_{xx}-\sigma_{yy})/\dot\gamma^2$ as a  
function of Peclet number Pe for three values of $\epsilon$. At low shear  
rates both $\eta$ and $\Psi_1$ saturate to a plateau value for fluid  
states but diverge in the glass. (Parameters: $\Gamma = 1, \gamma_c = 0.1, v_\sigma = 100, \nu = 1$.)
\label{fig1}
}
\end{figure}
\section{Results}
As mentioned above, in steady shear our tensorial schematic model gives for the flow curve $\sigma_{xy}(\dot\gamma)$ results similar to those of a model (there called the $F_{12}^{\dot\gamma}$ model) published previously \cite{faraday,fuchs3,fuchs4}. 
The latter model however says nothing about other elements of the stress tensor; 
nor nonsteady flows \cite{zausch}; nor non-shear flows. 
Below we give sample results from the new model that illuminate each of these 
issues in turn. 
Unless otherwise stated we select parameters $\Gamma = 1$, $\gamma_c = 0.1$, $v_\sigma = 100$, $\nu = 1$. 
The first of these choices, $\Gamma = 1$, sets the time unit; our choice directly equates strain rate to the Peclet number, Pe. (Continuing to assert this relation while choosing a different $\Gamma$ would allow flow curves to be scaled horizontally, 
possibly improving the fit to experiment.) 
The values chosen for $\gamma_c$ and $v_{\sigma}$ 
reflect the values seen experimentally for these two quantities 
($\sim 0.1, 100$, with the latter in units of $k_BT/d^3$) in hard sphere systems close to 
the glass transition \cite{flowcurveexperiments}. Note that smaller or larger values might be 
appropriate in systems with non hard-sphere interactions, such as short-range bonding forces, which, in 
the full model, enter  through changes in $S_k$. Turning to the mixing parameter $\nu$, our results 
for simple shear and planar extensional flows are independent of this parameter and for other flows 
display only a very weak $\nu$ dependence. 
For simplicity we therefore choose $\nu = 1$. 
The remaining variables are $\varepsilon$, the distance to the glass transition, and the 
flow history. 

\subsection{Normal stresses under steady shear}
The tensorial schematic model predicts the following diagonal 
stress tensor elements under steady shear flow
\begin{eqnarray}
\sigma_{xx}&=& 2\dot\gamma^2\,v_{\sigma}\int_{0}^{\infty} \!\!\!\!dt'
t'
\Phi^2(t') \,,
\label{diagonal_elements}
\\
\sigma_{yy}&=& 0 \hspace*{1cm}
\sigma_{zz}= 0,\notag
\end{eqnarray} 
where the correlator only depends on a single time argument due to restoration of stationarity.
Defining the first normal stress coefficient as $\Psi_1=(\sigma_{xx}-\sigma_{yy})/\dot\gamma^2=N_1/\dot\gamma^2$, we find
\begin{eqnarray}
\Psi_1= 2\,v_{\sigma}\int_{0}^{\infty} \!\!\!\!dt'
t'
\Phi^2(t').
\label{n1}
\end{eqnarray}
Here the correlator is the one calculated with the flow present, which decays to zero at long times. Thus $\Psi_1$ is finite and positive, meaning that in a steady shear experiment the bounding plates tend to get pushed apart. A positive $\Psi_1$ has been found recently in a full numerical solution of the MCT-ITT equations in 
two spatial dimensions \cite{oliver2D}, and the same is found for the full model in three dimensions if one invokes the isotropized (ISHSM) approximation.
On the other hand, the second normal stress difference $N_2=\sigma_{yy}-\sigma_{zz}$ is, from Eq.\ref{diagonal_elements}, identically zero, and this is also true of ISHSM. 

Fig.\ref{fig1} shows steady-state flow curves for the viscosity and the first normal stress coefficient in simple shear from the present model. 
In the fluid phase, $\varepsilon < 0$, both follow a power law as a function of the separation parameter, with $\eta(\dot\gamma\to 0) \sim (-\varepsilon)^{-2.34}$ and $\Psi_1(\dot\gamma \to 0) \sim (-\varepsilon)^{-4.68}$.
For $\varepsilon > 0$ (within the glass phase) both the viscosity and the first normal stress coefficient remain divergent as $\dot\gamma\to 0$, consistent with the presence of a dynamic yield stress -- that is, a finite limiting stress tensor at low shear rates. (As shown in \cite{faraday}, the full MCT-ITT predicts a nonzero dynamic yield stress throughout the glass phase; so does our schematic model as detailed below.)
As mentioned in the derivation of Eq.\ref{memory}, the viscosity curve almost coincides with the $F_{12}^{(\dot\gamma)}$ model of \cite{faraday}, which in turn closely approximates the full ISHSM predictions, and which themselves give a good account of experimental data \cite{fuchs3,flowcurveexperiments}. 
Note that, with the choice of units conventionally adopted in MCT-ITT, the Peclet number
in Fig.\ref{fig1} and subsequent figures coincides with the strain rate, Pe$=\dot\gamma d^2/D_0$ 
with $D_0$ the (bare) diffusivity.

\begin{figure}
\includegraphics[width=8cm]{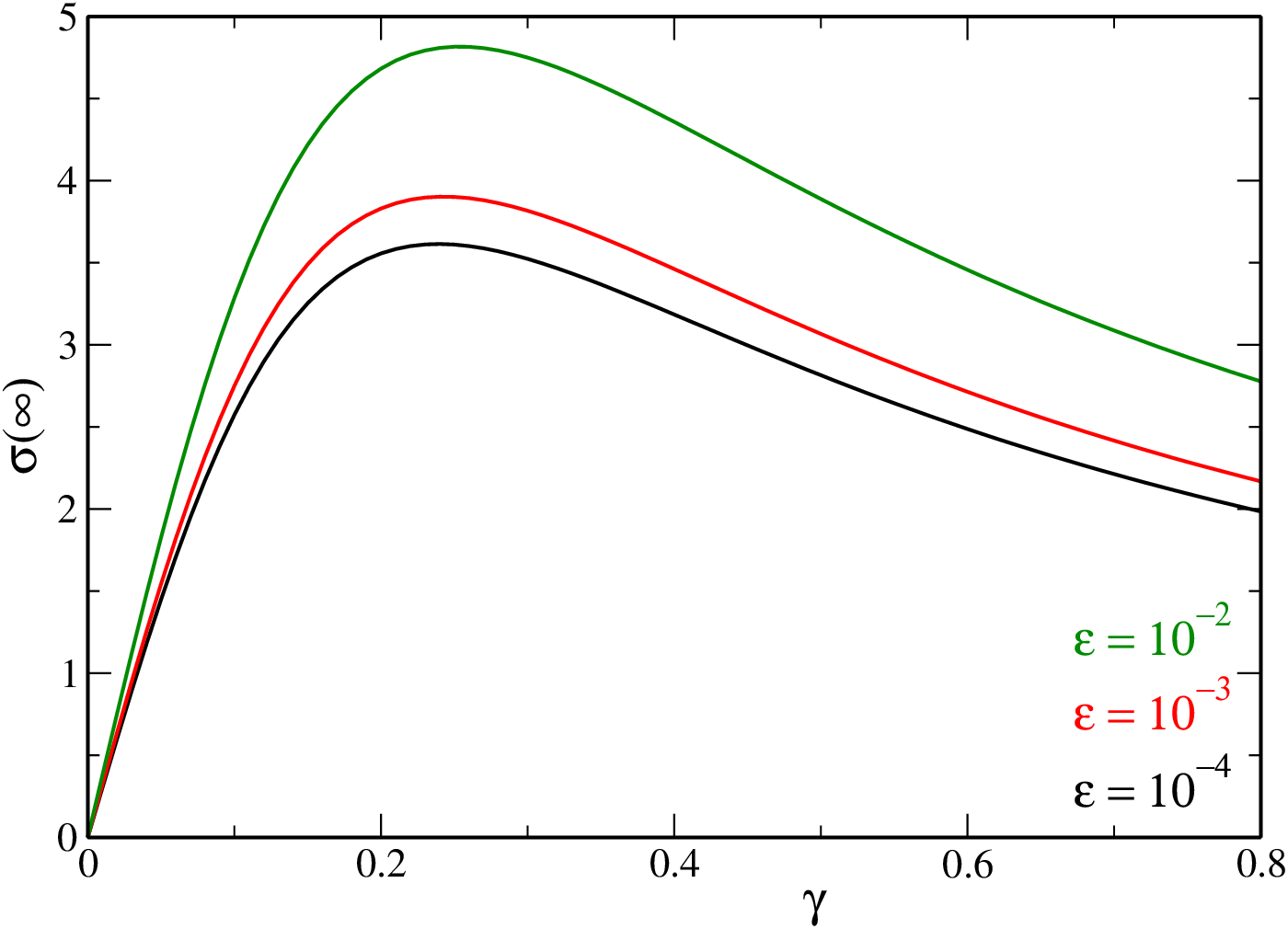}
\caption{
The long time shear stress following a step strain as a function of  
the step amplitude $\gamma$ for three glassy statepoints. At small  
values of $\gamma$ the expected linear response behavior is recovered.  
For larger $\gamma$ values a regime of sublinear increase indicates  
the onset of plastic flow. (Parameters: $\Gamma = 1, \gamma_c = 0.1, v_\sigma = 100, \nu = 1$.)
\vspace*{0.5cm}
\label{fig2}
}
\end{figure}
\subsection{Step shear strain}
In the case of a mathematically idealized step shear strain, 
$\dot\gamma(t)=\gamma\delta(t)$,  
Eq.\ref{schematicT2} simplifies considerably. (To attain this limiting case, one requires a strain ramp of duration $\tau\ll\Gamma^{-1}$, which may not be achievable experimentally. We leave the study of finite ramp rates to future work.)
The subsequent shear stress is given simply by
\begin{eqnarray}
\sigma_{xy}(t)=v_{\sigma}\gamma\Phi^2(t),
\label{step_stress}
\end{eqnarray}
where the correlator $\Phi(t)$ satisfies the linear equation
\begin{eqnarray}
\dot \Phi(t) 
&+& \Gamma\big(
\Phi(t) + 
\int_{0}^{t}dt'\, \tilde m(t-t')
\dot\Phi(t')
\big) =0,
\label{step_eom}\\
\notag 
\tilde m(t-t') &\equiv& 
\frac{\left(v_1\Phi_{\rm eq}(t-t')+v_2\Phi_{\rm eq}^2(t-t')\right)
}{1 + \left(\frac{\gamma}{2\gamma_c}\right)^2}\,.
\end{eqnarray}
Here $\Phi_{\rm eq}(t)$ is the equilibrium correlator obtained by solving Eq.\ref{eom2} 
in the absence of flow. The first normal stress difference 
$N_1=\sigma_{xx}-\sigma_{yy}$ for times following the step is given by
\begin{eqnarray}
N_1(t)=v_{\sigma}\gamma^2\Phi^2(t).
\label{step_n1}
\end{eqnarray}
This satisfies the 
Lodge-Meissner relationship $N_1(t)/\sigma_{xy}(t)=\gamma$ which holds for 
all constitutive models in the (misleadingly named) `simple fluids' class, of which ours is a member \cite{noll}.

The shear stress after step strain relaxes monotonically to zero in a fluid but has a nonzero
asymptote $\sigma(\infty)>0$ in the glass. Numerical calculations of $\sigma(\infty)$ using the
ISHSM \cite{brader1} showed this to be a nonmonotonic function of strain amplitude. This is a
consequence of the strain-erasure of memory which can over-compensate the (essentially linear)
dependence of the initial stress level on the strain, causing curves for different strain rates to cross. The physics of this effect is captured within our schematic model; Fig.\ref{fig2} shows $\sigma(\infty)$ as a function of $\gamma$ for various $\varepsilon$ and this is quite similar to Fig.1 of \cite{brader1}. There are deviations at large strain, caused by our choice of Eq.\ref{modified_h} to describe the strain effect. In fact, our schematic model may be more physical here than ISHSM which is found numerically to predict negative $\sigma(\infty)$ at large enough strains \cite{brader1}. (It is not yet clear whether the latter is an artefact of the isotropization used or signifies a deeper problem with MCT-ITT itself.)

\begin{figure}
\includegraphics[width=8cm]{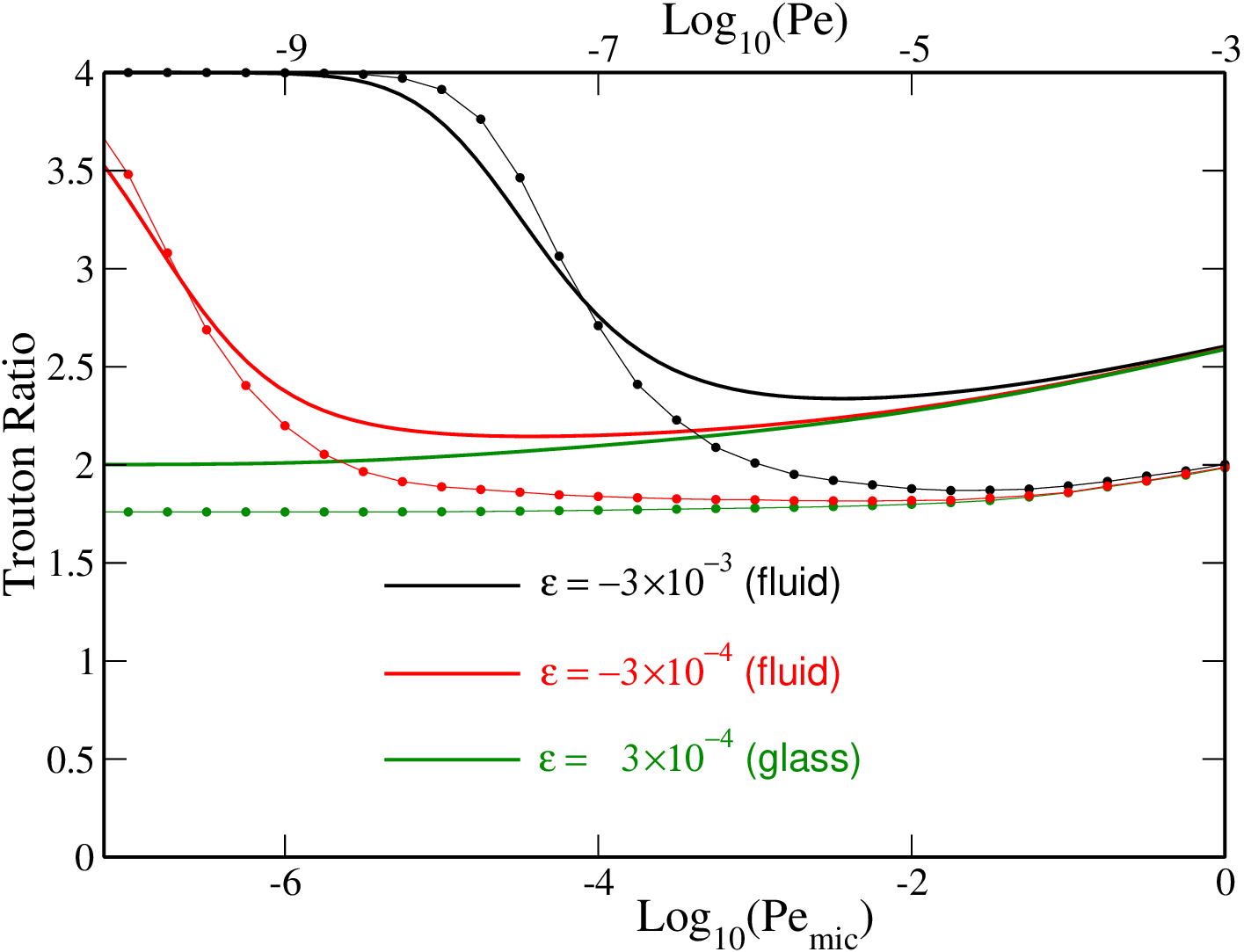}
\caption{The planar extensional Trouton ratio for two fluid statepoints  
($\varepsilon<0$) and one glassy state  
($\varepsilon>0$) calculated using the schematic model (full lines). 
For comparison we show also the microscopic results from \cite{brader2} (points connected by lines) 
where the full model was solved in the  isotropic (ISHSM) approximation. As the glass is approached 
the linear response regime vanishes, leading to a nontrivial Trouton ratio. 
The Peclet number employed in the microscopic calculations is denoted Pe$_{\rm mic}$. 
(Parameters: $\Gamma = 10^{-3}, \gamma_c = 0.1, v_\sigma = 100, \nu = 1$.)
\label{fig3}
}
\end{figure}
\subsection{Steady planar and uniaxial elongation}
For these flows, as in steady shear, our schematic model produces qualitatively similar flow curves to ISHSM. A sensitive test of the tensorial aspects of the model is to plot the ``Trouton ratio'', $(\sigma_{xx}-\sigma_{yy})/\sigma_{xy}$ as a function of 
Pe. Fig.\ref{fig3} shows a direct comparison
for the case of a planar extensional flow ($\kappa_{ij} = \dot\gamma(\delta_{xi}\delta_{xj}-\delta_{yi}\delta_{yj})$). For a fluid phase this must asymptote to $4$ at small strain rates, but in the glass a nontrivial (and smaller) value is possible. Within the schematic model, this nontrivial value is numerically indistinguishable from $2$, while the ISHSM gives a rather smaller number; for uniaxial extension the schematic model gives a noninteger value (close to $1.7$), this time slightly above ISHSM ($\simeq 1.6$).
A quantitative comparison of this kind requires a parameter-matching exercise to be undertaken. 
For this purpose, choosing $\varepsilon = 3(\phi-\phi_c)$ roughly matches the correlator's final relaxation time to a typical one in the ISHSM at volume fraction $\phi$ \cite{faraday}. To also match the rate of change of the yield stress with $\varepsilon$ to its ISHSM counterpart would require additional parameters in the schematic model. Qualitatively, this difference can be absorbed into an effective Pe number (equivalently, $\Gamma \neq 1$) as in Fig.\ref{fig3}.
Once this is done, the Trouton curves for schematic and ISHSM results are qualitatively very similar.

\subsection{Yield surface}
One striking feature of MCT-ITT is that it allows in principle an unambigous first-principles
determination of a dynamic yield stress for glasses. For any given flow geometry (e.g. simple shear,
planar elongation, or uniaxial elongation) a yield stress tensor is defined as the limiting stress
obtained as the relevant flow rate tends to zero from above. One expects that the manifold of these
yield stresses divides flowing states at large stress from non-flowing states at small stress and that this manifold forms a closed surface in some appropriately constructed space. 

We emphasize that the dynamic yield manifold may differ from that for static yield. The latter is
usually defined as the step stress that must be exceeded to result in steady flow; this may depend on
prior sample history whereas the dynamic yield manifold (being defined as the limiting stress within
a series of ergodic, fluidized steady states) does not. The relation between the two yield surfaces directly mirrors the familiar one between dynamic and static friction, and although they address different physical situations, each merits careful study. Here we present results only for the dynamic case which, because of the simpler flow history, is easier to compute. 
Crucially, however, our schematic model is tractable enough to allow prediction of 
the long-time response to step strains (or step stresses) of arbitrary character and amplitude 
(Fig.\ref{fig2}). 
Combining these, we may compile a conventional static yield manifold, and also address many other 
aspects of the flow-history-dependent yield response. 
We defer to future work an exploration of these problems, but believe that our model offers a promising new semiquantitative route, inspired by first principles statistical mechanics, into the study of diverse yield phenomena in amorphous solid materials, including static yield.

To calculate the dynamic yield manifold, we note first that for any particular flow geometry, the limiting stress tensor at vanishing flow rate can be diagonalized to give principal stresses $s_1,s_2,s_3$. Eliminating the (arbitrary) isotropic pressure for an incompressible system, we can then plot a point in the two dimensional $(s_1\!-\!s_2\,,\, s_2\!-\!s_3)$ plane. 
Repeating this procedure for all possible flows leads to a locus of points that define the `yield stress surface' in this two dimensional space. We call this `representation 1' of the yield manifold. In an alternative representation, all three stress variables are retained to give a surface in a three dimensional space; but since hydrostatic pressure is irrelevant (due to incompressibility) this surface must be translationally invariant along that axis. Thus looking down the hydrostatic axis again gives a closed curve (representation 2). 

\begin{figure}
\includegraphics[width=7cm]{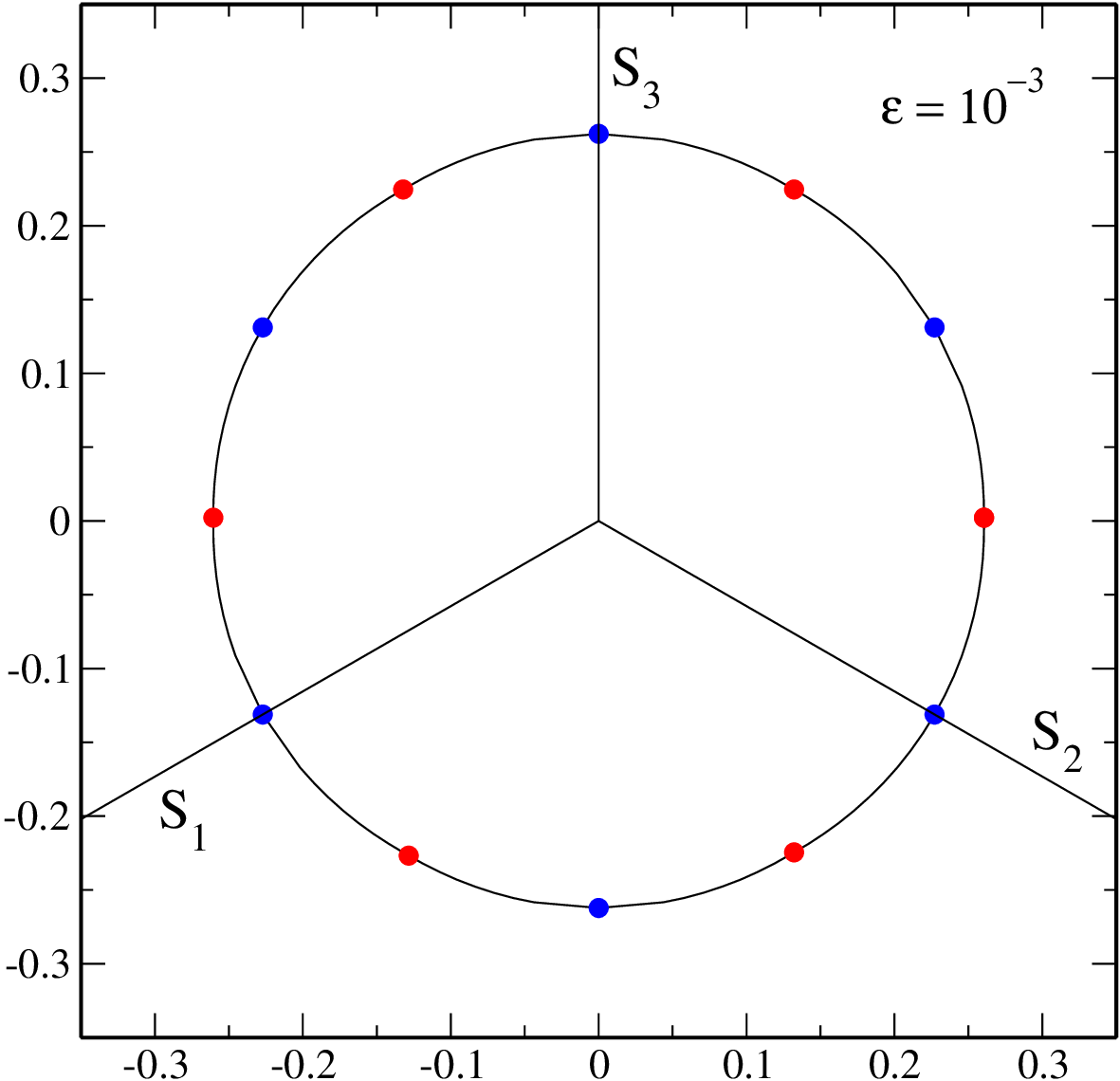}
\caption{
The dynamical yield surface for a glass with $\epsilon=10^{-3}$ as  
viewed along the hydrostatic axis $s_1=s_2=s_3$ in the space of  
principle stress differences. The red points correspond to planar  
extensional flow and the blue points to uniaxial extensional flow. The  
surface is not perfectly circular and exhibits maximal deviation from  
the von Mises circle at points of pure uniaxial extension. 
(Simple shear flows lie too close to planar extension to be shown separately.)
(Parameters: $\Gamma = 1, \gamma_c = 0.1, v_\sigma = 100, \nu = 1$.)
\label{fig4}
}
\end{figure}

Using symmetry and other arguments, it is possible to map out an entire yield surface by considering only a one-parameter family of flow geometries that interpolate between planar and uniaxial 
flows. (The explicit construction is given as Supplementary Information). The case of simple shear is not among this family of flows, but its stress is among the family of stresses spanned by the family, and shear flows in fact lie on the same yield surface to 
numerical accuracy.
The resulting yield surface is shown, in representation 2, in Fig.\ref{fig4}. This looks suggestively circular, but on close numerical study is found to deviate discernibly from a circle at the percent level, 
with the maximal deviation at the points of uniaxial extension. 
Our results for the yield surface are independent of $\Gamma$ and are insensitive to variations 
in $\nu$. Increases in $\epsilon$, $v_{\sigma}$ and $\gamma_c$ simply lead to a scaling of 
the surface, as the yield stress increases, e.g., on moving deeper into the glass.

Although ours is not a first-principles approach, but merely a schematic model inspired by one, we think the calculation of this yield surface  remarkable for two reasons. First, the yield surface in representation 2 is {\em almost} perfectly circular. Perfect circularity (corresponding to an ellipse in representation 1) is the content of `von Mises law', an empirical relation which has been used for nearly a century to approximate the yield surface of a wide range of materials \cite{hill}. The yield surfaces in question are static ones, whereas we calculate the dynamic: we therefore make no claim to have explained von Mises law, but do find the correspondence intriguing. It will be very interesting to see whether the static yield manifold computed from our model has a similar degree of circularity. In any case, since anisotropy must cause the yield condition to depend on material axes as well as principal stress axes, the starting point for any statistical-mechanical `derivation' of von Mises must presumably involve an isotropic (and therefore amorphous) solid -- that is, a glass. Our work represents a promising starting point for such an endeavor.

Secondly, the yield surface is not {\em quite} circular, despite all the simplifications that have been made in deriving the schematic model. 
\if{
Historically, empirical work on yield and plasticity has been guided first by considerations of simplicity, and then by the need to deviate from the simplest possible equations in order to accommodate an increasing body of experimental data \cite{hill}. As is evident historically from the development of constitutive models for polymer melts \cite{larson}, statistical mechanics offers an important alternative avenue to refinement, in which the improvements come from a closer alignment between empiricism and information derived from microscopic theory. 
}\fi
Indeed it can be shown for the schematic model that the circular yield manifold is recovered only under 
conditions where the first normal stress difference, at the point of yield under simple shear, is negligible (see Supplementary Information for details). 
This limit is approached whenever yield strains are small and might form the basis of a systematic
expansion with a von Mises-like circle as the zeroth order contribution. As found in experiments \cite{flowcurveexperiments} the yield strain for hard sphere colloids is indeed modest. 

\subsection{Conclusion}
We have presented a schematic model for the constitutive rheology of glasses, suppressing all wavevector indices but retaining much of the tensorial content of the full MCT-ITT approach \cite{catesfuchs,brader1,brader2}. The resulting tensorial structure of the schematic 
model satisfies applicable invariance laws for the nonlinear flow of materials without inertia \cite{noll}.
Our schematic model performs well at capturing qualitatively the behaviour of the full theory. More precisely, it gives results that are similar to those found by the (wavevector-dependent but still simplified) ISHSM approximation, in all flows for which results for the latter have so far been obtained. By construction, however, the schematic model is much easier to implement numerically across a wider range of flow geometries and histories. As a concrete example of this, we have calculated the full dynamic yield stress manifold by addressing a family of steady flows that interpolate between planar and uniaxial elongation. The resulting yield surface is very similar, but not quite identical, to the empirical form of von Mises that has been widely used to model (static, not dynamic) yield and plasticity in solids. It remains to be seen whether the static yield manifold for our schematic model is also of this form. More generally, our work offers promise for a better understanding of the physics of plasticity based on statistical mechanical principles applied to amorphous, isotropic solids -- specifically glasses -- and is a step towards the rational prediction of the general nonlinear rheology of this important group of materials.





\begin{acknowledgments}
We thank Oliver Henrich for valuable discussions, and the Transregio SFB TR6 and EPSRC/EP/E030173 for financial support. MEC holds a Royal Society Research Professorship. TV holds a Helmholtz Young Investigator Group fellowship (VH-NG 406).

\end{acknowledgments}

\section*{Appendix}
\subsection*{Relationship to Lodge and Maxwell equations}
The present schematic model is closely related to the Lodge equation \cite{larson}
\begin{eqnarray}
\sig(t) = \int_{-\infty}^{t}\!\!dt'\; \Finger(t,t') \;\frac{Ge^{-(t-t')/\lambda}}{\lambda},
\label{lodge1}
\end{eqnarray}
where $\lambda$ is the relaxation time of the system, $G$ is the instantaneous 
shear modulus and $\Finger(t,t')$ is the Finger tensor measuring the relative strain. 
The Lodge equation is valid in both the linear and nonlinear regimes and is a standard model 
in continuum rheological modelling, derivable from a number of simple molecular models, 
e.g. the dumbell model for dilute polymer solutions \cite{larson}.  
Integrating Eq.\ref{lodge1} by parts 
yields the alternative form
\begin{eqnarray}
\sig(t) =  
\int_{-\infty}^{t}\!\!dt'\; \left(-\frac{\partial}{\partial t'}
\Finger(t,t')\right) \;Ge^{-(t-t')/\lambda},
\label{lodge2}
\end{eqnarray}
which has a structure close to that of our schematic Eq.6. 
In this spirit we integrate Eq.6 by parts to obtain a Lodge-type expression
\begin{eqnarray}
\sig(t) = v_{\sigma}\int_{-\infty}^{t}\!\!dt'\; \Finger(t,t') 
\left( \frac{\partial}{\partial t'}\Phi^2(t,t') \right).
\label{parts}
\end{eqnarray}
The present theory thus goes considerably beyond the standard Lodge equation 
by incorporating memory which is both nonexponential and a function of two-time arguments, 
reflecting the lack of time-translational invariance under general flow.
Replacing the correlator with a simple exponential trivially recovers the Lodge equation.

The differential form of the Lodge equation is the upper-convected Maxwell equation \cite{larson}
, which is a simple nonlinear generalization of the familiar Maxwell model of 
viscoelasticity \cite{ll}. 
It is thus of interest to consider the differential form of Eq.6.  
Differentiation of Eq.\ref{parts} yields an integro-differential equation for the stress
\begin{eqnarray}
\frac{D\sig}{Dt} \;-\; 
v_{\sigma}\int_{-\infty}^{t}\!\!dt'\; \Finger(t,t') 
\left( \frac{\partial^2}{\partial t\,\partial t'}\Phi^2(t,t') \right) 
=2\Gamma\boldsymbol{\delta},
\label{integro_diff}
\end{eqnarray} 
where $\Gamma$ is the initial decay rate appearing in Eq.8, $\boldsymbol{\delta}$ is the unit tensor  
and we have introduced the upper-convected derivative \cite{hassager}
\begin{eqnarray}
\frac{D\sig}{Dt}=\dot{\sig}(t) - \kap(t)\,\sig(t) - \sig(t)\,\kap^{T}(t).
\label{upper}
\end{eqnarray}
If we neglect particle interactions and therefore memory effects in Eq.8 the correlator 
exhibits pure exponential decay, $\Phi(t,t')=\exp(-\Gamma(t-t'))$, 
and Eq.\ref{integro_diff} reduces to the upper-convected Maxwell equation \cite{larson}
\begin{eqnarray}
\lambda\frac{D\sig}{Dt} + \sig = \boldsymbol{\delta},
\label{upper_maxwell}
\end{eqnarray}
where $\lambda=1/2\Gamma$. 
The fact that Eq.\ref{integro_diff} cannot generally be reduced to simple 
differential form reflects the non-exponential character of the correlator. 
We note that the appearance of the upper-convected derivative in Eq.\ref{integro_diff} 
is due to the tensorial dependence on $\Finger(t,t')$ (a prefactor $\Finger^{-1}(t,t')$ 
would yield a lower-convected derivative).  

The Lodge equation is a member of the K-BKZ class of equations \cite{larson}, for which
\begin{eqnarray}
\sig(t)=2\int_{-\infty}^{t}\!\!dt'\;\left(
\frac{\partial u}{\partial I_1}\Finger(t,t') - \frac{\partial u}{\partial I_2}\Finger^{-1}(t,t')
\right).
\label{kbkz}
\end{eqnarray}
with $u(I_1,I_2,t-t')$ a function whose time integral gives the strain energy. 
This class of equations first arose in the context of rubber elasticity theory
and depending on the choice of Kernel $u$ admits a great variety of behaviour.
Even allowing for breakdown of time translation invariance, so that $u = u(I_1,I_2,t,t')$, we find that our model is not of this form (except for the limiting case of $\nu = 1$). However, modulo the same generalization of the time arguments, our model does belong to an extension of the Rivlin-Sawyers class of equations \cite{hassager} in which the coefficients of $\Finger$ and $\Finger^{-1}$ in Eq.\ref{kbkz} 
need not satisfy an integral relationship. 
This more general class of equations are capable of predicting a richer 
variety of rheological phenomena than that of the K-BKZ class. 
We note that the generalization to two time arguments in the strain energy kernel $u$ 
has nontrivial consequences and can give rise to a stress response significantly 
different from that of the standard K-BKZ theory (even in the case $\nu=1$).

\subsection*{The von Mises criterion}
The von Mises criterion is a well established empirical rule for determining the onset 
of plastic yield under an applied stress \cite{hill}. 
The criterion is typically expressed in terms of the principal stresses
\begin{eqnarray}		
\frac{1}{6}\left( (s_1-s_2)^2 + 
(s_2-s_3)^2 + (s_1-s_3)^2\right)= (\sigma_{ss}^{y})^2\,,
\label{vonmises_yield}
\end{eqnarray}
where $\sigma_{ss}^{y}$ is the shear stress at yield under simple shear deformation. 
Eq.\ref{vonmises_yield} is based on the physical assumption that yielding occurs 
when the distortion strain energy (defined as the left hand side of Eq.\ref{vonmises_yield} 
divided by $2G$, where G is the shear modulus) exceeds a critical value. 
It is useful to interpret Eq.\ref{vonmises_yield} geometrically in the space of 
principal stresses, where it describes a surface separating elastically deformed states 
from states of plastic flow. 
Eq.\ref{vonmises_yield} defines a circular cylinder with axis along the line 
$s_1=s_2=s_3$ and radius $\sqrt{2}\,\sigma_{ss}^{y}$. 
The invariance along the line $s_1=s_2=s_3$ is a geometrical reflection 
of the fact that the yield condition is independent of hydrostatic pressure.

\subsection*{Flow parameterization}
For calculation of the dynamic yield stress surface we employ a one-parameter family of 
flow geometries which interpolate between planar and uniaxial flow. 
Consider the following parameterized Finger tensor
\begin{eqnarray}            
     \Finger=\left(
     \begin{array}{ccc}
     e^{2\gamma} & 0 & 0 \\
     0 & e^{-(1+A)\gamma} & 0 \\
     0 & 0 & e^{-(1-A)\gamma} 
     \end{array}   
     \right).   
\label{parameter_finger}           
\end{eqnarray}
By varying the parameter $A$ from $0$ to $1$ the flow changes continuously from uniaxial to planar 
elongation. 
Using our schematic model to calculate the dynamic yield stress for each value of $A$ in 
this range enables one twelfth of the yield surface to be completed. 
General symmetry requirements then enable the yield surface to be completed without further 
calculation \cite{hill}. 
The flows characterized by Eq.\ref{parameter_finger} for $0<A<1$ are thus 
physically distinct. 
Other cases, such as uni-biaxial flow ($A<0$), are degenerate in the sense that they do not provide 
additional information regarding the yield surface.

Substitution of Eq.\ref{parameter_finger} into Eq.6 yields the principal stresses
\begin{eqnarray}
s_1&=& v_{\sigma}\dot\gamma \int_{0}^{\infty}dt'\, 
\left(2e^{2\dot\gamma t'} \right)\Phi^2(t'),
\notag\\
s_2&=& v_{\sigma}\dot\gamma \int_{0}^{\infty}dt'\,\left( 
-(1+A)\,e^{-(1+A)\dot\gamma t'} 
\right)\Phi^2(t'),
\notag\\
s_3&=& v_{\sigma}\dot\gamma \int_{0}^{\infty}dt'\,\left( 
-(1-A)\,e^{-(1-A)\dot\gamma t'} 
\right)\Phi^2(t'),
\label{vonmises_proof1}
\end{eqnarray} 
where the correlators are those calculated using Eqs.8-10 with the flow given by 
Eq.\ref{parameter_finger}. 
Connection to the von Mises criterion can be made by
expanding the exponentials in Eq.\ref{vonmises_proof1} to first order in $\dot\gamma t$. 
This yields a moment expansion
\begin{eqnarray}\label{vonmises_proof2}
\frac{s_1}{v_{\sigma}}\!\!\!\!&=&\!\!\!\! 2\dot\gamma\!\! \int_{0}^{\infty}\!\!dt'\,\Phi^2(t')
+
4\dot\gamma^2 \!\!\int_{0}^{\infty}\!\!dt'\,t'\Phi^2(t'),
\\
\frac{s_2}{v_{\sigma}}\!\!\!\!&=&\!\!\!\! -(A+1)\dot\gamma \!\!\int_{0}^{\infty}\!\!dt'\,\Phi^2(t')
+
(A+1)^2\dot\gamma^2 \!\!\int_{0}^{\infty}\!\!dt'\,t'\Phi^2(t'),
\notag\\
\frac{s_3}{v_{\sigma}}\!\!\!\!&=&\!\!\!\! (A-1)\dot\gamma \!\!\int_{0}^{\infty}\!\!dt'\,\Phi^2(t')
+
(A-1)^2\dot\gamma^2 \!\!\int_{0}^{\infty}\!\!dt'\,t'\Phi^2(t').\notag
\end{eqnarray} 
Defining the integral 
\begin{eqnarray}
J_n = v_{\sigma}\dot\gamma^{n+1} \int_{0}^{\infty}\!\!dt'\,(t')^{n}\Phi^2(t'),
\label{integrals}
\end{eqnarray}
and substituting Eqs.\ref{vonmises_proof2} into the left hand side 
of Eq.\ref{vonmises_yield} yields
\begin{eqnarray}	
\frac{1}{6}\left( (s_1-s_2)^2 + 
(s_2-s_3)^2 + (s_1-s_3)^2\right)&=& \notag\\ 
&&\hspace*{-6cm}(3+A^2)J_0^2+\frac{1}{3}(3+A^2)^2J_1^2 + 6(1-A^2)J_1J_2.
\label{vonmises_yield1}	
\end{eqnarray} 
Using Eqs.8 and 9 it can be shown that in the limit of small flow rate 
$\dot\gamma\rightarrow 0$
\begin{eqnarray}
J_0 &&\!\!\!\!\!\rightarrow \frac{v_{\sigma}}{\sqrt{3+A^2}}\,\dot\gamma
\int_{0}^{\infty}\!\!dt'\,\Phi_{{\rm ss},\dot\gamma\rightarrow 0}^2(t')
= \frac{\sigma_{\rm ss}^y}{\sqrt{3+A^2}}, 
\notag\\
J_1 &&\!\!\!\!\!\rightarrow \frac{v_{\sigma}}{3+A^2}\,\dot\gamma^2
\int_{0}^{\infty}\!\!dt'\,t'\Phi_{{\rm ss},\dot\gamma\rightarrow 0}^2(t')
= \frac{N_{1}^y}{2(3+A^2)}, 
\label{j0j1}
\end{eqnarray} 
where $N_1^y$ is the first normal stress 
difference at yield under steady simple shear.
Substitution of Eq.\ref{j0j1} into Eq.\ref{vonmises_proof2} thus yields
\begin{eqnarray}\label{vonmises_yield2}		
\frac{1}{6}\left( (s_1-s_2)^2 + 
(s_2-s_3)^2 + (s_1-s_3)^2\right)&=& \notag\\
&&\hspace*{-5.4cm}(\sigma_{\rm ss}^y)^2 + \frac{1}{12}(N_1^y)^2
+\frac{3(1-A^2)}{(3+A^2)^{3/2}}N_1^y\sigma_{\rm ss}^y.
\end{eqnarray} 
This expression is the dynamic analogue of the von Mises criterion for static 
yield. 
Neglecting the first normal stress difference reduces Eq.\ref{vonmises_yield2} to the von 
Mises form. 
Numerical calculations using the schematic model show that $N_1^y$ 
is approximately two orders of magnitude smaller than $\sigma_{\rm ss}^y$ for a given 
value of $\epsilon$. 
The third term on the right hand side of 
Eq.\ref{vonmises_yield2} thus constitutes a small correction to circularity. 
The dependence of the third term on the parameter $A$ lends non-trivial structure to 
the yield surface. 
Continuing the moment expansion (Eq.\ref{vonmises_proof2}) to higher orders in $\dot\gamma t$ 
generates further small corrections and may thus be viewed as a systematic 
expansion about a circular dynamic yield surface.
The numerical results presented in the main text were performed using Eq.\ref{vonmises_proof1} 
without further approximation.


$$\,$$

\end{document}